# A feature-based framework for detecting technical outliers in water-quality data from in situ sensors

**Priyanga Dilini Talagala**
Department of Econometrics and Business Statistics, Monash University, Australia, and
ARC Centre of Excellence for Mathematics and Statistical Frontiers
Email: dilini.talagala@monash.edu
Corresponding author

**Rob J. Hyndman**
Department of Econometrics and Business Statistics, Monash University, Australia, and
ARC Centre of Excellence for Mathematics and Statistical Frontiers

**Catherine Leigh**
Institute for Future Environments, Science and Engineering Faculty, Queensland University of Technology, Australia, and
ARC Centre of Excellence for Mathematics and Statistical Frontiers

**Kerrie Mengersen**
Science and Engineering Faculty, Queensland University of Technology, Australia, and
ARC Centre of Excellence for Mathematics and Statistical Frontiers

**Kate Smith-Miles**
School of Mathematics and Statistics, University of Melbourne, Australia, and
ARC Centre of Excellence for Mathematics and Statistical Frontiers

# A feature-based framework for detecting technical outliers in water-quality data from in situ sensors

**Abstract**

Outliers due to technical errors in water-quality data from *in situ* sensors can reduce data quality and have a direct impact on inference drawn from subsequent data analysis. However, outlier detection through manual monitoring is unfeasible given the volume and velocity of data the sensors produce. Here, we proposed an automated framework that provides early detection of outliers in water-quality data from *in situ* sensors caused by technical issues. The framework was used first to identify the data features that differentiate outlying instances from typical behaviours. Then statistical transformations were applied to make the outlying instances stand out in transformed data space. Unsupervised outlier scoring techniques were then applied to the transformed data space and an approach based on extreme value theory was used to calculate a threshold for each potential outlier. Using two data sets obtained from *in situ* sensors in rivers flowing into the Great Barrier Reef lagoon, Australia, we showed that the proposed framework successfully identified outliers involving abrupt changes in turbidity, conductivity and river level, including sudden spikes, sudden isolated drops and level shifts, while maintaining very low false detection rates. We implemented this framework in the open source R package `oddwater`.

## 1 Introduction

Water-quality monitoring traditionally relies on water samples collected manually. The samples are then analyzed within laboratories to determine the water-quality variables of interest. This type of rigorous laboratory analysis of field-collected samples is crucial in making natural resources management decisions that affect human welfare and environmental conditions. However, with the rapid advances in hardware technology, the use of *in situ* water-quality sensors positioned at different geographic sites is becoming an increasingly common practice used to acquire real-time measurements of environmental and water-quality variables. Though only a subset of the required water-quality variables can be measured by these sensors, they have several advantages. Their ability to collect large quantities of data and to archive historic records allows for deeper analysis of water-quality variables to improve understanding about field conditions and water-quality processes (Glasgow et al. 2004). Near-real-time monitoring also allows operators to identify and respond to potential issues quickly and thus manage the operations efficiently. Further, the use of *in situ* sensors can greatly reduce the labor involved in field sampling and laboratory analysis.

Water-quality sensors are exposed to changing environments and extreme weather conditions, and thus are prone to errors, including failure. Automated detection of outliers in water-quality data from *in situ* sensors has therefore captured the attention of many researchers both in the ecology and data science communities (Hill, Minsker & Amir 2009; Archer, Baptista & Leen 2003; Raciti, Cucurull & Nadjm-Tehrani 2012; McKenna et al. 2007; Koch & McKenna 2010). This problem of outlier detection in water-quality data from *in situ* sensors can be divided into two sub-topics according to their focus: (1) identifying errors in the data due to issues unrelated to water events per se, such as technical aberrations, that make the data unreliable and untrustworthy; and (2) identifying real events (e.g. rare but sudden spikes in turbidity associated with rare but sudden high-flow events). Both problems are equally important when making natural resource management decisions that affect human welfare and environmental conditions. Problem 1 can also be considered as a data preprocessing phase before addressing Problem 2. Most of the outlier detection techniques presented in the literature focus on detecting real events (McKenna et al. 2007; Koch & McKenna 2010; Raciti, Cucurull & Nadjm-Tehrani 2012), however, the characterization of outliers as technical errors has been understudied (Chen, Kher & Somani 2006; Sharma, Golubchik & Govindan 2010). For example, Shahid, Naqvi & Qaisar (2015) highlight the importance of detecting technical errors while emphasizing the lack of research attention given to the topic.

In this work we focus on Problem 1, i.e. detecting unusual measurements caused by technical errors that make data unreliable and untrustworthy, and affect performance of any subsequent data analysis under Problem 2. According to Yu (2012), the degree of confidence in the sensor data is one of the main requirements for a properly defined environmental analysis procedure. For instance, researchers and policy makers are unable to use water-quality data containing technical outliers with confidence for decision making and reporting purposes, because erroneous conclusions regarding the quality of the water being monitored could ensue, leading, for example, to inappropriate or unnecessary water treatment, land management or warning alerts to the public (Kotamäki et al. 2009; Rangeti et al. 2015). Missing values and corrupted data can also have an adverse impact on water-quality model building and calibration processes (Archer,

Baptista & Leen 2003). Early detection of these technical outliers will limit the use of corrupted data for subsequent analysis. For instance, it will limit the use of corrupted data in real-time forecasting and online applications such as on-line drinking water-quality monitoring and early warning systems (Storey, Van der Gaag & Burns 2011), predicting algal bloom outbreaks leading to fish kill events and potential human health impacts, forecasting water level and currents etc. (Glasgow et al. 2004; Archer, Baptista & Leen 2003; Hill & Minsker 2006). However, because data arrive near continuously at high speed in large quantities, manual monitoring is highly unlikely to be able to capture all the errors. These issues have therefore increased the importance of developing automated methods for early detection of outliers in water-quality data from *in situ* sensors (Hill, Minsker & Amir 2009).

Different statistical approaches are available to detect outliers in water-quality data from *in situ* sensors. For example, Hill & Minsker (2006) addressed the problem of outlier detection in environmental sensors using regression-based time series models. In this work they addressed the scenario as a univariate problem. Their prediction models are based on four data-driven methods: naive, clustering, perceptron, and Artificial Neural Networks (ANN). Measurements that fell outside the bounds of an established prediction interval were declared as outliers. They also considered two strategies: anomaly detection (AD) and anomaly detection and mitigation (ADAM) for the detection process. ADAM replaces detected outliers with the predicted value prior to the next predictions whereas AD simply uses the previous measurements without making any alteration to the detected outliers. These types of data-driven methods develop models using sets of training examples containing a feature set and a target output. Later, Hill, Minsker & Amir (2009) addressed the problem by developing three automated anomaly detection methods using dynamic Bayesian networks (DBN) and showed that DBN-based detectors using either robust Kalman filtering or Rao-Blackwellized particle filtering, outperformed that of Kalman filtering.

Another common approach for detecting outliers in environmental sensor data is based on residuals (the differences between predicted and actual values). Due to the ability of ANNs to model a wide range of complex non-linear phenomena, Moatar, Fessant & Poirel (1999) used ANN techniques to detect anomalies such as abnormal values, discontinuities, and drifts in pH readings. After developing the pH model, the Student t-test and the cumulative Page–Hinkley test were applied to detect changes in the mean of the residuals to detect measurement error occurring over short periods of time. The work was later expanded to a multivariate scenario with some additional water-quality variables including dissolved oxygen, electrical conductivity, pH and temperature (Moatar, Miquel & Poirel 2001). Their proposed algorithm used both deterministic and stochastic approaches for the model building process. Observed data were then compared with the model forecasts using a set of classical statistical tests to detect outliers, demonstrating the effectiveness and advantages of the multimodel approach. Later, Archer, Baptista & Leen (2003) proposed a method to detect failures in the water-quality sensors due to biofouling based on a sequential likelihood ratio test. Their method also had the ability to provide estimates of biofouling onset time, which was useful for the subsequent step of outlier correction.

A common feature of all of the above methods is that they are usually employed in a supervised or semi-supervised context and thus require training data pre-labeled with known outliers or data that are free from the anomalous features of interest. In many cases, however, not all the possible outliers are known in advance and can arise spontaneously as new outlying behaviors during the test phase. In such situations, supervised methods may fail to detect those outliers. Semi-supervised methods are also unsuitable for certain applications due to the unavailability of training data containing only typical instances that are free from outliers (Goldstein & Uchida 2016). The data sets that we consider in this paper suffer from both of these limitations highlighting the need for a more general approach.

Our key objective of this research was therefore to propose an unsupervised framework to detect technical outliers in high frequency water-quality data measured by *in situ* sensors. In this work we mainly focus on outliers involving abrupt changes in value, including sudden spikes, sudden isolated drops and level shifts. First, we propose an unsupervised framework that provides early detection of technical outliers in water-quality data from *in situ* sensors. Rule-based methods were also incorporated into the proposed framework to flag occurrence of impossible, out-of-range, and missing values. Second, we provide a comparative analysis of the efficacy and reliability of both density- and nearest neighbor distance-based outlier scoring techniques. Third, we introduce an R (R Core Team 2018) package, `oddwater` (Talagala & Hyndman 2018) that implements the proposed framework and related functions.

Our proposed framework has many advantages: (1) it can take the correlation structure of the water-quality variables into account when detecting outliers; (2) it can be applied to both univariate and multivariate problems; (3) the outlier scoring techniques that we consider are unsupervised, data-driven approaches and therefore do not require training data sets for the model building process, and can be extended easily to other time series from other sites; (4) the outlier thresholds have a probabilistic interpretation as they are based on extreme value theory; (5) the framework can easily be extended to streaming data such that it can provide near-real-time support; and (6) the proposed framework has the ability to deal with irregular (unevenly spaced) time series.

## 2 Materials and Methods

Our proposed framework for detecting outliers in water-quality data from *in situ* sensors has six main steps (Figure 1), and the structure of this section is organised accordingly.

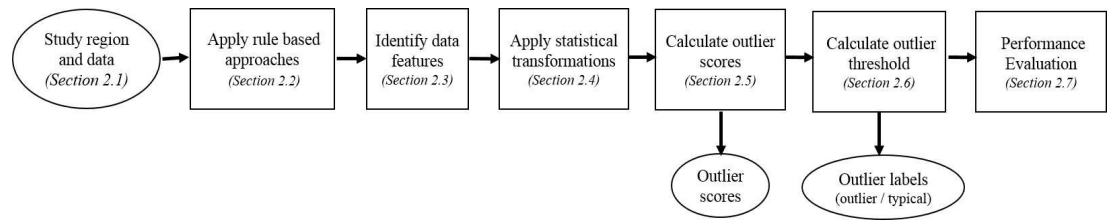

**Figure 1:** *The proposed framework for outlier detection in water quality data from in situ sensors. Squares represents the main steps involved. Circles correspond to input and output.*

### 2.1 Study region and data

To evaluate the effectiveness of our proposed framework we considered a challenging real-world problem of monitoring water-quality using *in situ* sensors in a natural river system. This is challenging because the system is susceptible to a wide range of environmental, biological and human impacts that can lead to variation in water-quality and affect the technological performance of the sensors. For comparison, we evaluated two study sites, Sandy Creek and Pioneer River (PR), both in the Mackay-Whitsunday region of northeastern Australia (Mitchell, Brodie & White 2005). These two rivers flow into the Great Barrier Reef lagoon, and have catchment areas of 1466 $km^2$ and 326 $km^2$, respectively. In this region, the wet season typically occurs from December to April and is dominated by higher rainfall and air temperatures, whereas the dry season typically occurs from May to November with lower rainfall and air temperatures (McInnes et al. 2015). The sensors at these two sites are housed within a monitoring station on the river banks. Water is pumped from the rivers to the stations approximately every 60 or 90 minutes to take measurements of various water-quality variables that are logged by the sensors. Here we focused on three water-quality variables: turbidity (NTU), conductivity ($\mu S/cm$) and river level ($m$).

The water-quality data obtained from *in situ* sensors located at Sandy Creek were available from 12 March 2017 to 12 March 2018. The data set included 5402 recorded points. These time series were irregular (i.e. the frequency of observations was not constant) with a minimum time gap of 10 minutes and a maximum time gap of around 4 hours. The data obtained from Pioneer River were available from 12 March 2017 to 12 March 2018, and included 6280 recorded points. Many missing values were observed during the initial part of all three series, turbidity, conductivity and river level, at Pioneer River. With the help of water-quality experts, observations were labeled as outliers or not, with the aim of evaluating the performance of the proposed framework.

### 2.2 Apply rule-based approaches

Following Thottan & Ji (2003), we incorporated simple rules into our outlier detection framework to detect outliers such as out-of-range values, impossible values (e.g. negative values) and missing values, and labeled and removed them prior to applying the statistical transformations introduced in Section 2.4.

If a sensor reading was outside the corresponding sensor detection range it was marked as an outlier. Negative readings are also inaccurate and impossible for river turbidity, conductivity and level. We therefore imposed a simple constraint on the algorithm to filter these values and mark them as outliers. Missing values are also frequently encountered in water-quality sensor data (Rangeti et al. 2015). We detected missing values by calculating the time gaps between readings. If a gap exceeded the maximum allowable time difference between any two consecutive readings, the corresponding time stamp was then marked as an outlier due to missingness. Here, the maximum allowable time difference was set at 180 minutes, given that the water-quality measurements were set to be taken at least every 90 minutes.

### 2.3 Identify data features

After removal of out-of-range, impossible and missing values, we then identified common characteristics or patterns of the possible types of outliers in water-quality data that would differentiate them from typical instances or events. For turbidity, for example, "extreme" deviations upward are more likely than deviations downwards (Panguluri et al. 2009). The opposite is true for conductivity (Tutmez, Hatipoglu & Kaymak 2006). Further, in a turbidity time series a sudden isolated upward shift (spike) is a point outlier (a single observation that is surprisingly large or small, independent of the neighboring observations (Goldstein & Uchida 2016)), but if the sudden upward shift is followed by a gradually decaying tail then it becomes part of the typical behavior. For river level, rates of rise are often fast compared with fall rates. In general, isolated data points that are outside the general trend are outliers. Further, natural water processes under typical conditions generally tend to be comparatively slow; sudden changes therefore mostly correspond to outlying behaviors. Hereafter, these characteristics will be referred to as 'data features'.

## 2.4 Apply statistical transformations

After identifying the data features, different statistical transformations were applied to the time series to highlight different types of outliers, focusing on sudden isolated spikes, sudden isolated drops, sudden shifts, and clusters of spikes (Table 1) that deviate from the typical characteristics of each variable (Leigh et al. 2018).

**Table 1:** *Transformation methods used to highlight different types of outliers in water-quality sensor data. Let $Y_t$ represent an original series from one of the three variables: turbidity, conductivity and level at time t.*

| Transformation | Formula | Data Feature | Focus |
| --- | --- | --- | --- |
| Log transformation | $\log(y_t)$ | High variability of the data. | To stabilize the variance across time series and make the patterns more visible (e.g. level shifts) |
| First difference | $\log(y_t/y_{t-1})$ | Isolated spikes (in both positive and negative directions) that are outside the general trend are considered as outliers. Under typical behavior, sudden upward (downward) shifts are possible for turbidity (conductivity), but their rate of fall (rise) is generally slower than the rate of rise (fall). | To separate isolated spikes from the general upward/downward trend patterns. |
| Time gap | $\Delta t$ | | To identify missing values. |
| First derivative | $x_t = \log(y_t/y_{t-1})/\Delta t$ | Data are unevenly-spaced time series. | To handle irregular time series. Data points with large gaps will get small value. Large gaps indicate the lack of information to make a claim regarding the points. |
| One sided derivative | | | |
| *Turbidity or level* | $\min\{x_t, 0\}$ | Extreme upward trend in turbidity and level under typical behavior. | To separate spikes from typical upward trends. |
| *Conductivity* | $\max\{x_t, 0\}$ | Extreme downward trend in conductivity under typical behavior. | To separate isolated drops from typical downward trends. |
| Rate of change | $(y_t - y_{t-1})/y_t$ | High or low variability in the data. | To detect change points in variance. |
| Relative difference | $y_t - (1/2)(y_{t-1} + y_{t+1})$ | Natural processes are comparatively slow. Sudden changes (upward or downward movements) typically correspond to outlying instances. | To detect sudden changes (both upward and downward movements) |

In this work we considered the outlier detection problem in the multivariate setting. By applying different transformations on water-quality variables we converted our original problem of outlier detection in the temporal context to a non-temporal context through a high dimensional data space (Figure 2). Different transformations were applied on both horizontal and vertical axes resulting in different data patterns. We evaluated the performance of the transformations (Dang & Wilkinson 2014) using the maximum separability of the two classes: outliers and typical points. For example, in the data obtained from Sandy Creek, the one sided derivative transformation clearly separated most of the target outlying points from the typical points (Figure 2(e)).

When the transformation involves both the current value $Y_t$ and the lagged value $Y_{t-1}$ (as in the first difference, first derivative, and one sided derivatives), the neighboring points can emerge as outliers instead of the actual outlying point. For an example, if an outlier occurs at time point $t$, then the two values derived from the first derivative transformation ($(y_t - y_{t-1})$ and $(y_{t+1} - y_t)$) get highlighted as outlying values because they both involve $y_t$. That is each outlying instance is now represented by two consecutive values under the first derivative transformation. The goal of the one sided derivative transformation is to filter one high value for each outlying instance. However the high values obtained could correspond to either the actual outlying time point or the neighboring time point, because each transformed value is derived from two consecutive observations. If the primary focus of detecting technical outliers is to alert managers of sensor failures, then it will be inconsequential if the alarm is triggered either at the actual time point corresponding to the outlier or at the next immediate time point. However if the purpose is different, such as producing a trustworthy dataset by labeling or correcting detected outliers, then additional conditions should be imposed to ensure that the time points declared as outliers correspond to the actual outlying points and not to their immediate neighboring points.

## 2.5 Calculate outlier scores

We considered eight unsupervised outlier scoring techniques for high dimensional data, involving nearest neighbor distances or densities of the observations. Methods based on $k$-nearest neighbor distances (where $k \in Z^+$) were the HDoutliers algorithm (Wilkinson 2018), KNN-AGG and KNN-SUM algorithms (Angiulli & Pizzuti 2002; Madsen 2018) and Local Distance-based Outlier Factor (LDOF) algorithm (Zhang, Hutter & Jin 2009), which calculate the outlier score under the assumption that any outlying point (or outlying clusters of points) in the data space is (are) isolated; therefore the outliers are those points having the largest $k$-nearest neighbor distances. In contrast, the density based Local Outlier Factor (LOF) (Breunig et al. 2000), Connectivity-based Outlier Factor (COF) (Tang et al. 2002), Influenced Outlierness (INFLO) (Jin et al. 2006) and Robust Kernel-based Outlier Factor (RKOF) (Gao et al. 2011) algorithms calculate an outlier score based on how isolated a point is with respect to its surrounding neighbors, and therefore the outliers are those points having the lowest densities (see Supplementary materials for detail). Each algorithm assigns outlier scores for all the data points in the high dimensional space that described the degree of outlierness of the individual data points such that outliers are those points having the largest scores (Kriegel, Kröger & Zimek 2010; Shahid, Naqvi & Qaisar 2015). This step allowed us to set a domain specific threshold (Section 2.6) to select the most relevant outliers (Chandola, Banerjee & Kumar 2009).

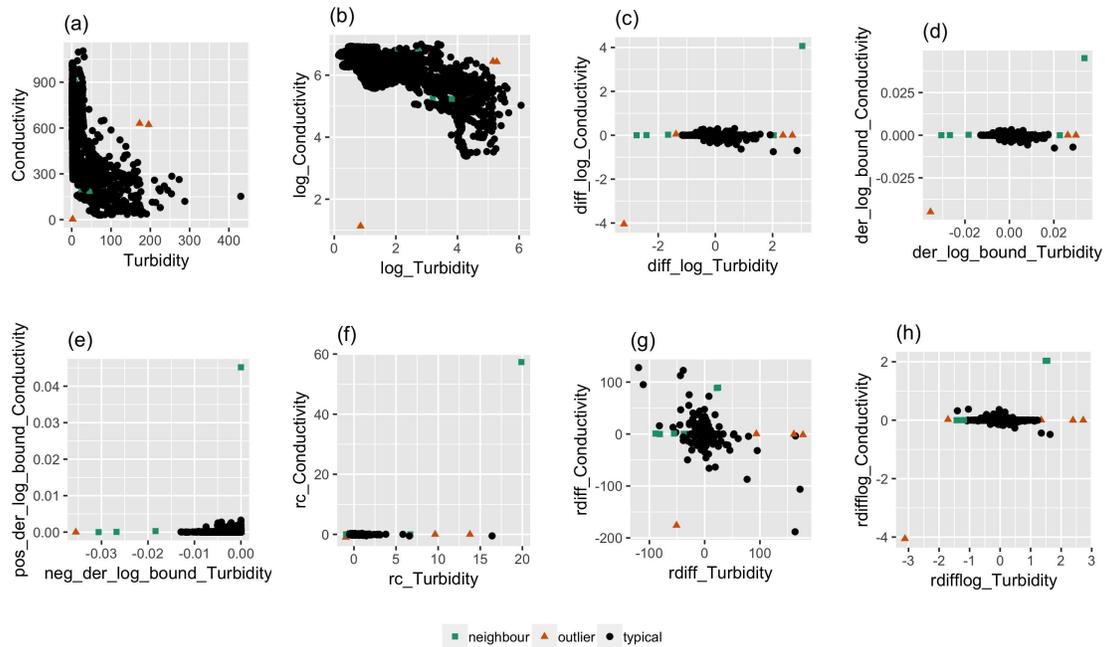

**Figure 2:** *Bivariate relationships between transformed series of turbidity and conductivity measured by in situ sensors at Sandy Creek. In each scatter plot, outliers determined by water-quality experts are shown in red, while typical points are shown in black. (a) Original series, (b) Log transformation, (c) First difference, (d) First derivative, (e) One sided derivative, and (f) Rate of change, (g) Relative difference (for original series), (h) Relative difference (for log transformed series).*

## 2.6 Calculate outlier threshold

Following Schwarz (2008), Burridge & Taylor (2006) and Wilkinson (2018), we used extreme value theory (EVT) to calculate an outlier threshold and assign a bivariate label for each point either as an outlier or typical point. The threshold calculation process started from a subset of data containing the 50% of observations with the smallest outlier scores, under the assumption that this subset contained the outlier scores corresponding to typical data points and the remaining subset contained the scores corresponding to the possible candidates for outliers. Following Weissman's spacing theorem (Weissman 1978), the algorithm then fit an exponential distribution to the upper tail of the outlier scores of the first subset, and computed the upper $1 - \alpha$ (in this work $\alpha$ was set to 0.05) points of the fitted cumulative distribution function, thereby defining an outlying threshold for the next outlier score. From the remaining subset the algorithm then selected the point with the smallest outlier score. If this outlier score exceeded the cutoff point, all the points in the remaining subset were flagged as outliers and searching for outliers ceased. Otherwise, the point was declared as a non-outlier and was added to the subset of the typical points. The threshold was then updated by including the latest addition. The searching algorithm continued until an outlier score was found that exceeded the latest threshold (Schwarz 2008). We performed this threshold calculation under the assumption that the distribution of outlier scores produced by each of the eight unsupervised outlier scoring techniques for high dimensional data was in the maximum domain of attraction of the Gumbel distribution which consists of distribution functions with exponentially decaying tails including the exponential, gamma, normal and log-normal (Embrechts, Klüppelberg & Mikosch 2013).

## 2.7 Performance evaluation

We performed an experimental evaluation on the accuracy and computational efficiency of the proposed framework with respect to the eight outlier scoring techniques, using the different transformations and different combinations of variables (turbidity, conductivity and river level) (Table 1). These experimental combinations were evaluated with respect to common measures for binary classification based on the values of the confusion matrix which summarizes the false positives (FP; i.e. when a typical observation is misclassified as an outlier), false negatives (FN; i.e. when an actual outlier is misclassified as a typical observation), true positives (TP; i.e. when an actual outlier is correctly classified), and true negatives (TN; i.e. when an observation is correctly classified as a typical point). The measures include accuracy (($TP + TN$)/($TP + FP + FN + TN$)) which explains the overall effectiveness of a classifier; error rate (ER = ($FP + FN$)/($TP + FP + FN + TN$)) which explains the misclassification of the classifier; and geometric-mean (GM = $\sqrt{TP * TN}$) which explains the relative balance of TP and TN of the classifier (Sokolova & Lapalme 2009). According to Hossin & Sulaiman (2015), these measures are not enough to capture the poor performance of the classifiers in the presence of imbalanced data sets where the size of the typical class (positive class) is much larger than the outlying class (negative class). The data sets obtained from *in situ* sensors were highly imbalanced and negatively dependent (i.e. containing many more typical observations than outliers). Therefore, we used three additional measures that are recommended for imbalanced problems with only two classes (i.e. typical and outlying) by Ranawana & Palade (2006): the negative predictive value ($NPV = TN/(FN + TN)$) which measures the probability of a negatively predicted pattern actually being negative; positive predictive value ($PPV = TP/(TP + FP)$) which measures the probability of a positively predicted pattern actually being positive; and optimized precision (OP) which is a combination of accuracy, sensitivity and specificity metrics (Ranawana & Palade 2006).

To evaluate the performance of our proposed framework we incorporated additional steps after detecting the outlying time points using the outlying threshold based on EVT. This was done because the time points declared as outliers by the outlying threshold could correspond to either the actual outlying points or to their neighbors. Once the time points were declared as outliers, the corresponding points in the high dimensional space were further investigated by comparing their positions with respect to the median of the typical points declared by our proposed framework. This step allowed us to find the most influential variable for each outlying point. For example, in Figure 2(e) the isolated point in the first quadrant is an outlier in the two dimensional space due to the outlying behavior of the conductivity measurement. In contrast, the four isolated points in the third quadrant are outliers due to the outlying behavior of the turbidity measurement. After detecting the most influential variables for each outlying instance, further investigations were carried out separately for each individual outlying instance to see whether the outlying instance was due to a sudden spike or a sudden drop by comparing the direction of the detected points with respect to the mean of its two immediate surrounded neighbors and itself. These additional steps in the algorithm allowed us to make the necessary corrections if the neighboring points were declared as outliers instead of the actual outliers.

Using the outlier threshold, our proposed framework assigns a bivariate label (either as outlier or typical point) to each observed time point and thereby creates a vector of predicted class labels. That is, if a time point is declared as an outlier by the proposed algorithm, then that could be due to at least one variable in the data set. We also declared each time point as an outlier or not based on the labels assigned by the water quality experts. At a given time point, if at least one variable was labeled as an outlier by the water quality experts then the corresponding time point was marked as an outlier and thereby creating a vector of ground-truth labels. Then the performance measures were calculated based on these two vectors of ground-truth labels and predicted class labels.

### 2.8 Software implementation

The proposed framework was implemented in the open source R package `oddwater` (Talagala & Hyndman 2018), which provides a growing list of transformation and outlier scoring methods for high dimensional data together with visualization and performance evaluation techniques. Version 0.5.0 of the package was used for the results presented herein and is available from Github (github.com/pridiltal/oddwater). The datasets used for this article are also available in the package `oddwater`. We measured the computation time (minimum ($min_t$), mean ($mu_t$), maximum ($max_t$) execution time) using the `microbenchmark` package (Mersmann 2018) for different combinations of algorithms, transformations and variable combinations on 28 core Xeon-E5-2680-v4 @ 2.40GHz servers.

## 3 Results

### 3.1 Analysis of water-quality data from *in situ* sensors at Sandy Creek

A negative relationship was clearly visible between the water-quality variables: turbidity and conductivity and also between conductivity and river level measured by *in situ* sensors at Sandy Creek (Figures 3 and 4(a,c)). Further, no clear separation was observed between the target outliers and the typical points in the original data space (Figure 4(a–c)). However, a clear separation was apparent between the two sets of points once the one sided derivative transformation (an appropriate transformation for unevenly spaced data) was applied to the original series (Figures 4(d–f) and 5 ).

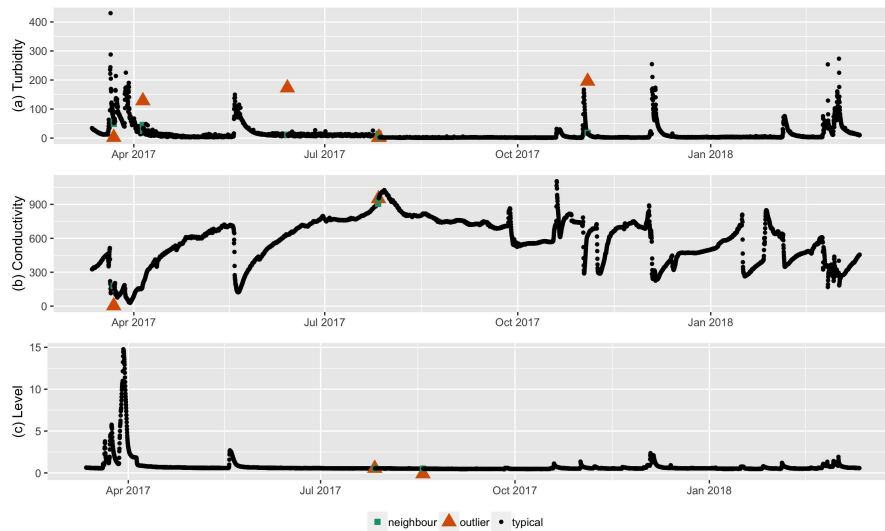

**Figure 3:** *Time series for turbidity (NTU), conductivity (μS/cm) and river level (m) measured by in situ sensors at Sandy Creek. In each plot, outliers determined by water-quality experts are shown in red. Typical points are shown in black.*

KNN-AGG and KNN-SUM algorithms performed on all three water-quality variables together, and on turbidity and conductivity together using the one sided derivative transformation, gave the highest OP (0.8329) and NPV values(0.9996), which are the most recommended measurements for negatively dependent data where the focus is more on sensitivity (the proportion of positive patterns being correctly recognized as being positive) than specificity (Ranawana & Palade 2006).

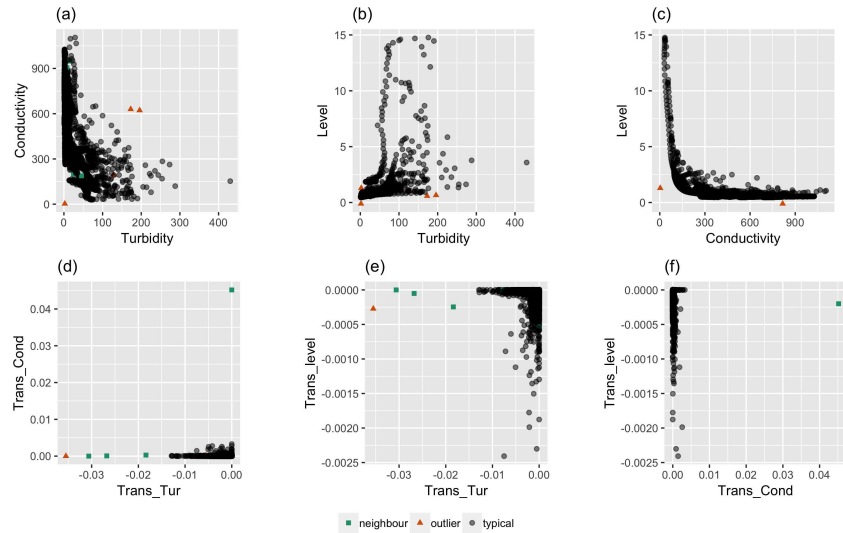

**Figure 4:** *Top panel (a–c): Bi-variate relationships between original water-quality variables (turbidity (NTU), conductivity (µS/cm) and river level (m)) measured by in situ sensors at Sandy Creek. Bottom panel (d–f): Bi-variate relationships between transformed series (one sided derivative) of turbidity (NTU), conductivity (µS/cm) and river level (m) measured by in situ sensors at Sandy Creek. In each scatter plot, outliers determined by water-quality experts are shown in red, while typical points are shown in black. Neighboring points are marked in green*

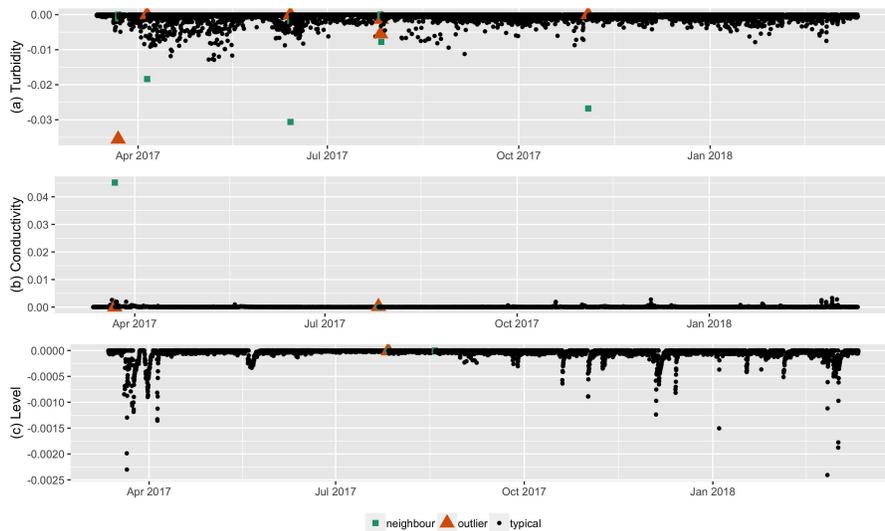

**Figure 5:** *Transformed series (one sided derivatives) of turbidity (NTU), conductivity (µS/cm) and river level (m) measured by in situ sensors at Sandy Creek. In each plot outliers determined by water-quality experts are shown in red, while typical points are shown in black.*

**Table 2:** *Performance metrics of outlier detection algorithms performed on multivariate water-quality time series data (T, turbidity; C, conductivity; L, river level) from in situ sensors at Sandy Creek, arranged in descending order of OP values. See Sections 2.7-8 for performance metric codes and details.*

| i | Variables | Transformation | Method | TN | FN | FP | TP | Accuracy | ER | GM | OP | PPV | NPV | min_t | mu_t | max_t |
|---|---|---|---|---|---|---|---|---|---|---|---|---|---|---|---|---|
| 1 | T-C-L | One sided Derivative | KNN-AGG | 5394 | 2 | 1 | 5 | 0.9994 | 0.0006 | 164.2255 | 0.8329 | 0.8333 | 0.9996 | 378.51 | 404.01 | 493.41 |
| 2 | T-C-L | One sided Derivative | KNN-SUM | 5394 | 2 | 1 | 5 | 0.9994 | 0.0006 | 164.2255 | 0.8329 | 0.8333 | 0.9996 | 177.16 | 186.85 | 270.29 |
| 3 | T-C | One sided Derivative | HDoutliers | 5396 | 2 | 0 | 4 | 0.9996 | 0.0004 | 146.9149 | 0.7996 | 1.0000 | 0.9996 | 102.70 | 112.94 | 195.38 |
| 4 | T-C | One sided Derivative | KNN-AGG | 5395 | 2 | 1 | 4 | 0.9994 | 0.0006 | 146.9013 | 0.7995 | 0.8000 | 0.9996 | 381.82 | 411.65 | 517.97 |
| 5 | T-C | One sided Derivative | KNN-SUM | 5395 | 2 | 1 | 4 | 0.9994 | 0.0006 | 146.9013 | 0.7995 | 0.8000 | 0.9996 | 177.90 | 190.39 | 286.17 |
| 6 | T-C | First Derivative | HDoutliers | 5393 | 2 | 3 | 4 | 0.9991 | 0.0009 | 146.8741 | 0.7993 | 0.5714 | 0.9996 | 40.54 | 45.02 | 72.08 |
| 7 | T-C | First Derivative | KNN-AGG | 5392 | 2 | 4 | 4 | 0.9989 | 0.0011 | 146.8605 | 0.7992 | 0.5000 | 0.9996 | 386.26 | 415.80 | 489.43 |
| 8 | T-C-L | First Derivative | KNN-AGG | 5395 | 4 | 0 | 3 | 0.9993 | 0.0007 | 127.2203 | 0.5993 | 1.0000 | 0.9993 | 377.63 | 404.37 | 476.17 |
| 9 | T-C-L | First Derivative | KNN-SUM | 5395 | 4 | 0 | 3 | 0.9993 | 0.0007 | 127.2203 | 0.5993 | 1.0000 | 0.9993 | 179.24 | 188.91 | 273.28 |
| 10 | T-C | First Derivative | KNN-SUM | 5396 | 4 | 0 | 2 | 0.9993 | 0.0007 | 103.8846 | 0.4993 | 1.0000 | 0.9993 | 178.99 | 189.52 | 283.51 |
| 11 | T-C | First Derivative | LDOF | 5395 | 4 | 1 | 2 | 0.9991 | 0.0009 | 103.8749 | 0.4991 | 0.6667 | 0.9993 | 17261.49 | 17444.70 | 17809.90 |
| 12 | T-C | One sided Derivative | LDOF | 5395 | 4 | 1 | 2 | 0.9991 | 0.0009 | 103.8749 | 0.4991 | 0.6667 | 0.9993 | 17024.26 | 17253.76 | 18079.39 |
| 13 | T-C-L | First Derivative | HDoutliers | 5395 | 5 | 0 | 2 | 0.9991 | 0.0009 | 103.8749 | 0.4435 | 1.0000 | 0.9991 | 48.65 | 52.48 | 66.94 |
| 14 | T-C-L | One sided Derivative | HDoutliers | 5395 | 5 | 0 | 2 | 0.9991 | 0.0009 | 103.8749 | 0.4435 | 1.0000 | 0.9991 | 110.43 | 118.21 | 193.23 |
| 15 | T-C-L | Original series | KNN-AGG | 5394 | 5 | 1 | 2 | 0.9989 | 0.0011 | 103.8653 | 0.4434 | 0.6667 | 0.9991 | 376.61 | 391.65 | 465.28 |
| 16 | T-C-L | First Derivative | COF | 5393 | 5 | 2 | 2 | 0.9987 | 0.0013 | 103.8557 | 0.4433 | 0.5000 | 0.9991 | 5869.09 | 5939.83 | 6394.22 |
| 17 | T-C-L | One sided Derivative | COF | 5393 | 5 | 2 | 2 | 0.9987 | 0.0013 | 103.8557 | 0.4433 | 0.5000 | 0.9991 | 5676.82 | 5787.44 | 6238.38 |
| 18 | T-C-L | Original series | LDOF | 5393 | 5 | 2 | 2 | 0.9987 | 0.0013 | 103.8557 | 0.4433 | 0.5000 | 0.9991 | 17078.47 | 17156.85 | 17308.09 |
| 19 | T-C-L | One sided Derivative | INFLO | 5392 | 5 | 3 | 2 | 0.9985 | 0.0015 | 103.8460 | 0.4432 | 0.4000 | 0.9991 | 1071.47 | 1113.64 | 1177.61 |
| 20 | T-C-L | One sided Derivative | LDOF | 5392 | 5 | 3 | 2 | 0.9985 | 0.0015 | 103.8460 | 0.4432 | 0.4000 | 0.9991 | 17181.45 | 17261.90 | 17435.77 |
| 21 | T-C-L | One sided Derivative | LOF | 5392 | 5 | 3 | 2 | 0.9985 | 0.0015 | 103.8460 | 0.4432 | 0.4000 | 0.9991 | 500.21 | 516.91 | 596.91 |
| 22 | T-C-L | Original series | RKOF | 5392 | 5 | 3 | 2 | 0.9985 | 0.0015 | 103.8460 | 0.4432 | 0.4000 | 0.9991 | 322.26 | 353.95 | 426.41 |
| 23 | T-C-L | One sided Derivative | RKOF | 5387 | 5 | 8 | 2 | 0.9976 | 0.0024 | 103.7979 | 0.4426 | 0.2000 | 0.9991 | 338.87 | 370.48 | 464.05 |
| 24 | T-C-L | Original series | INFLO | 5386 | 5 | 9 | 2 | 0.9974 | 0.0026 | 103.7882 | 0.4424 | 0.1818 | 0.9991 | 1034.30 | 1070.70 | 1136.71 |
| 25 | T-C-L | First Derivative | INFLO | 5381 | 5 | 14 | 2 | 0.9965 | 0.0035 | 103.7401 | 0.4418 | 0.1250 | 0.9991 | 1076.54 | 1107.92 | 1168.28 |
| 26 | T-C-L | First Derivative | RKOF | 5380 | 5 | 15 | 2 | 0.9963 | 0.0037 | 103.7304 | 0.4417 | 0.1176 | 0.9991 | 341.62 | 369.74 | 456.11 |
| 27 | T-C | First Derivative | COF | 5396 | 5 | 0 | 1 | 0.9991 | 0.0009 | 73.4575 | 0.2848 | 1.0000 | 0.9991 | 5881.54 | 5991.83 | 6552.44 |
| 28 | T-C | Original series | KNN-AGG | 5395 | 5 | 1 | 1 | 0.9989 | 0.0011 | 73.4507 | 0.2846 | 0.5000 | 0.9991 | 371.54 | 405.05 | 483.66 |
| 29 | T-C | First Derivative | LOF | 5394 | 5 | 2 | 1 | 0.9987 | 0.0013 | 73.4439 | 0.2845 | 0.3333 | 0.9991 | 498.26 | 512.32 | 596.10 |
| 30 | T-C | One sided Derivative | INFLO | 5394 | 5 | 2 | 1 | 0.9987 | 0.0013 | 73.4439 | 0.2845 | 0.3333 | 0.9991 | 1153.12 | 1207.01 | 1281.94 |
| 31 | T-C | One sided Derivative | COF | 5394 | 5 | 2 | 1 | 0.9987 | 0.0013 | 73.4439 | 0.2845 | 0.3333 | 0.9991 | 5755.00 | 5880.81 | 6420.84 |
| 32 | T-C | Original series | LDOF | 5394 | 5 | 2 | 1 | 0.9987 | 0.0013 | 73.4439 | 0.2845 | 0.3333 | 0.9991 | 16842.06 | 17022.86 | 17414.38 |
| 33 | T-C | Original series | RKOF | 5393 | 5 | 3 | 1 | 0.9985 | 0.0015 | 73.4370 | 0.2844 | 0.2500 | 0.9991 | 321.05 | 351.80 | 440.30 |
| 34 | T-C | First Derivative | INFLO | 5392 | 5 | 4 | 1 | 0.9983 | 0.0017 | 73.4302 | 0.2842 | 0.2000 | 0.9991 | 1134.59 | 1194.86 | 1271.15 |
| 35 | T-C | First Derivative | RKOF | 5392 | 5 | 4 | 1 | 0.9983 | 0.0017 | 73.4302 | 0.2842 | 0.2000 | 0.9991 | 335.07 | 363.16 | 435.72 |
| 36 | T-C | Original series | INFLO | 5387 | 5 | 9 | 1 | 0.9974 | 0.0026 | 73.3962 | 0.2835 | 0.1000 | 0.9991 | 1095.21 | 1143.56 | 1219.37 |
| 37 | T-C | One sided Derivative | LOF | 5384 | 5 | 12 | 1 | 0.9969 | 0.0031 | 73.3757 | 0.2831 | 0.0769 | 0.9991 | 501.13 | 511.33 | 585.68 |
| 38 | T-C | One sided Derivative | RKOF | 5380 | 5 | 16 | 1 | 0.9961 | 0.0039 | 73.3485 | 0.2826 | 0.0588 | 0.9991 | 339.51 | 368.29 | 456.57 |
| 39 | T-C-L | First Derivative | LDOF | 5395 | 6 | 0 | 1 | 0.9989 | 0.0011 | 73.4507 | 0.2489 | 1.0000 | 0.9989 | 17253.87 | 17323.23 | 17400.86 |
| 40 | T-C-L | First Derivative | LOF | 5395 | 6 | 0 | 1 | 0.9989 | 0.0011 | 73.4507 | 0.2489 | 1.0000 | 0.9989 | 504.55 | 517.12 | 604.40 |
| 41 | T-C-L | Original series | KNN-SUM | 5395 | 6 | 0 | 1 | 0.9989 | 0.0011 | 73.4507 | 0.2489 | 1.0000 | 0.9989 | 164.66 | 177.26 | 243.39 |
| 42 | T-C-L | Original series | COF | 5395 | 6 | 0 | 1 | 0.9989 | 0.0011 | 73.4507 | 0.2489 | 1.0000 | 0.9989 | 5864.44 | 5931.75 | 6329.68 |
| 43 | T-C-L | Original series | LOF | 5395 | 6 | 0 | 1 | 0.9989 | 0.0011 | 73.4507 | 0.2489 | 1.0000 | 0.9989 | 480.25 | 504.97 | 576.20 |
| 44 | T-C-L | Original series | HDoutliers | 5394 | 6 | 1 | 1 | 0.9987 | 0.0013 | 73.4439 | 0.2487 | 0.5000 | 0.9989 | 45.39 | 48.60 | 60.32 |
| 45 | T-C | Original series | KNN-SUM | 5396 | 6 | 0 | 0 | 0.9989 | 0.0011 | 0.0000 | -0.0011 | NaN | 0.9989 | 172.67 | 184.63 | 272.46 |
| 46 | T-C | Original series | COF | 5396 | 6 | 0 | 0 | 0.9989 | 0.0011 | 0.0000 | -0.0011 | NaN | 0.9989 | 5826.28 | 5896.44 | 6804.28 |
| 47 | T-C | Original series | LOF | 5396 | 6 | 0 | 0 | 0.9989 | 0.0011 | 0.0000 | -0.0011 | NaN | 0.9989 | 477.04 | 502.67 | 567.97 |
| 48 | T-C | Original series | HDoutliers | 5395 | 6 | 1 | 0 | 0.9987 | 0.0013 | 0.0000 | -0.0013 | 0.0000 | 0.9989 | 38.11 | 41.75 | 66.34 |

Based on OP values, the one sided derivative transformation outperformed the first derivative transformation (Table 2, rows 1–5 compared to rows 6–10). Further, the distance-based outlier detection algorithms HDoutliers, KNN-AGG and KNN-SUM outperformed all others (Table 2, rows 1–10 compared to rows 11–48). Among the three methods the performance of *k*-nearest neighbor distance-based algorithms were only slightly higher (OP = 0.8329) than the HDoutliers algorithm (OP= 0.7996), which is based only on the nearest neighbor distance. The algorithm combinations with the five highest OP values also had high PPV (approximately 0.8). Furthermore, considering river level for the detection of outliers in the water-quality sensors slightly improved the performance (OP = 0.8329). Among the analysis with transformed series, LOF with the first derivative transformation performed the least well (OP= 0.2489). For the most of the outlier detection algorithms (KNN-SUM, KNN-AGG, HDoutliers, COF, LOF and INFLO) the poorest performances were associated with the untransformed original series, having the lowest OP and NPV values, highlighting how data transformation can improve the ability of outlier detection algorithms while maintaining low false detection rates.

The three outlier detection algorithms that demonstrated the highest level of accuracy (HDoutliers, KNN-AGG and KNN-SUM) also outperformed the others with respect to computational time. HDoutliers algorithm required the least computational time. Among the remaining two, the mean computational time of KNN-AGG ($\approx$ 400 milliseconds) was twice that of KNN-SUM's (< 200 milliseconds). LOF and its extensions (INFLO, COF and LDOF) demonstrated the poorest performance with respect computational time (> 500) milliseconds on average).

Only KNN-SUM and KNN-AGG assigned high scores to most of the targeted outliers in turbidity, conductivity and level data transformed using the one-sided derivative (Figure 6(a,b)). For each outlying instance however, the next immediate neighboring point was assigned the high outlier score instead of the true outlying point. After determining the most influential variable using the additional steps of the proposed algorithm (Section 2.7), adjustments were made to correct this to the actual outlier. The outlier scores produced by LOF and COF (Figure 6(d,f)) were unable to capture the outlying behaviors correctly and demonstrated high scattering. With respect to both accuracy and computational efficiency, the KNN-SUM algorithm maintained the greatest balance between false positive and false negative rates (Figure 7).

### 3.2 Analysis of water-quality data from *in situ* sensors at Pioneer River

Some of the target outliers in the data obtained from the *in situ* sensors at Pioneer River only deviated slightly from the general trend (Figure 8), making outlier detection challenging. A negative relationship was clearly visible between turbidity and conductivity (Figure 9(a)), however the relationship between level and conductivity was complex (Figure 9(c)). Most of the target outliers were masked by the typical points in the original space (Figure 9(a–c)). Similar to Sandy Creek, data obtained from the sensors at Pioneer River showed good separation between outliers and typical points under the one sided derivative transformation (Figures 9(d–f) and 10). However, the sudden spikes in turbidity labeled as outliers by water-quality experts could not be separated from the majority by a large distance and were only visible as a small group (micro cluster (Goldstein & Uchida 2016)) in the boundary defined by the typical points (Figure 9(d, e)).

From the performance analysis it was observed that turbidity and conductivity together produced better results (Table 3, rows 1–3) than when combined with level, which tended to reduce the performance (i.e. generating lower OP and NPV values) while increasing the false negative rate (Table 3, rows 4–5). KNN-AGG and KNN-SUM (Table 3, rows 1–3) had the highest accuracy

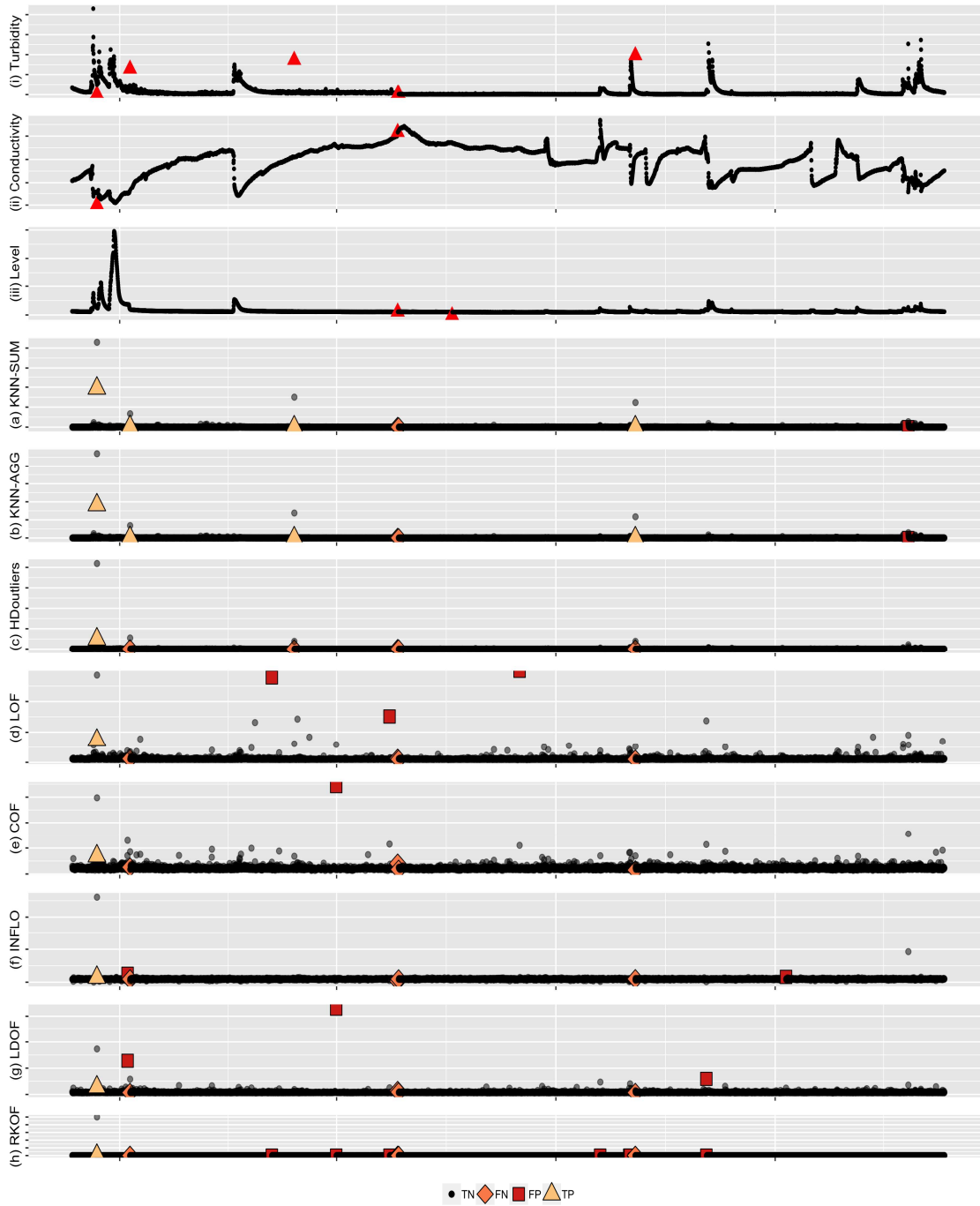

**Figure 6:** *Classification of outlier scores produced from different algorithms as true negatives (TN), true positives (TP), false negatives (FN), false positives (FP). The top three panels (i, ii, iii) correspond to the original series (turbidity, conductivity and river level) measured by in situ sensors at Sandy Creek. The target outliers (detected by water-quality experts) are shown in red, while typical points are shown in black. The remaining panels (a–h) give outlier scores produced by different outlier detection algorithms for high dimensional data when applied to the transformed series (one sided derivative) of the three variables: turbidity, conductivity and level.*

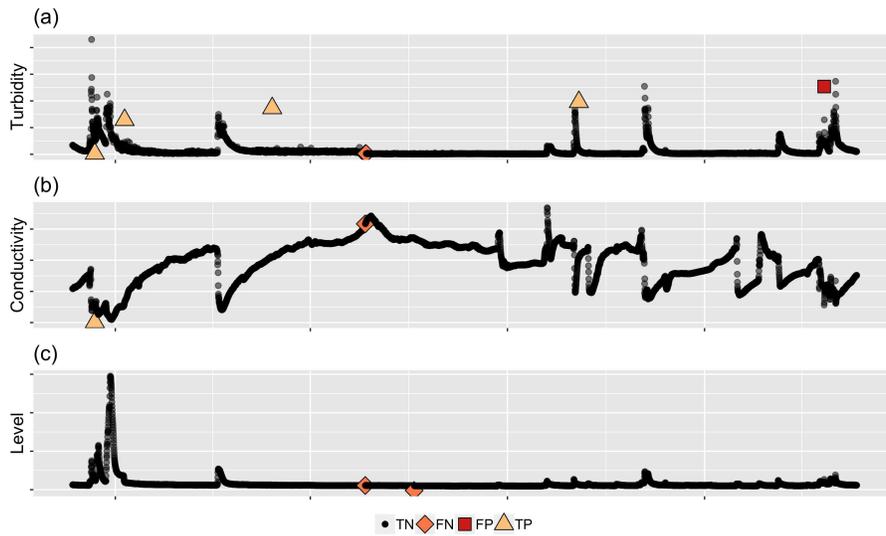

**Figure 7:** *Classification of turbidity (T), conductivity (C), level (L) observations measured by in situ sensors at Sandy Creek by KNN-SUM algorithm as true negatives (TN), true positives (TP), false negatives (FN), false positives (FP) when applied to the TCL combination with one sided derivative transformation.*

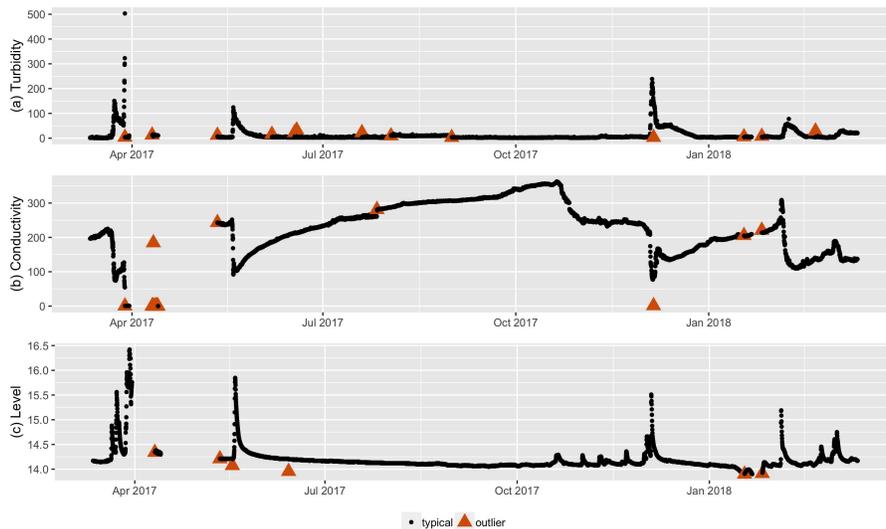

**Figure 8:** *Time series for turbidity (NTU), conductivity (µS/cm) and river level (m) measured by in situ sensors at Pioneer River. In each plot, outliers determined by water-quality experts are shown in red, while typical points are shown in black.*

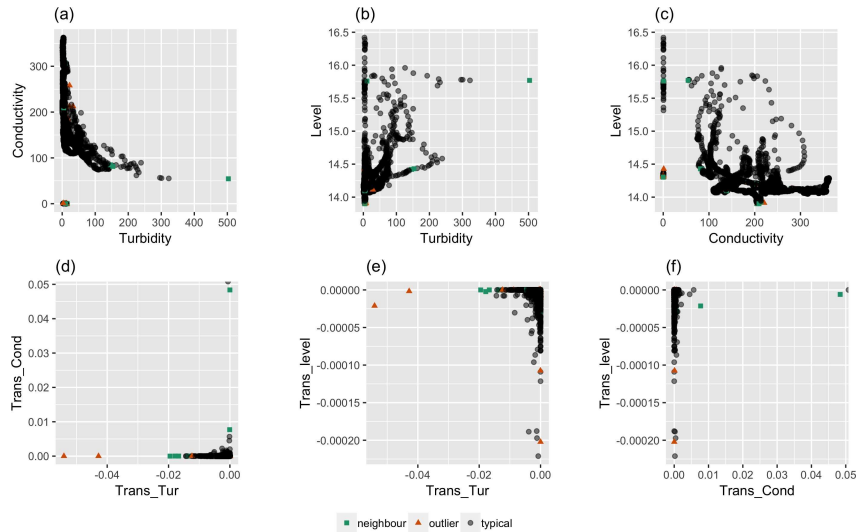

**Figure 9:** *Top panel (a–c): Bi-variate relationships between original water-quality variables (turbidity (NTU), conductivity (μS/cm) and river level (m)) measured by in situ sensors at Pioneer River. Bottom panel (d–f): Bi-variate relationships between transformed series (one sided derivative) of turbidity (NTU), conductivity (μS/cm) and river level (m) measured by in situ sensors at Pioneer River. In each scatter plot, outliers determined by water-quality experts are shown in red, while typical points are shown in black. Neighboring points are marked in green.*

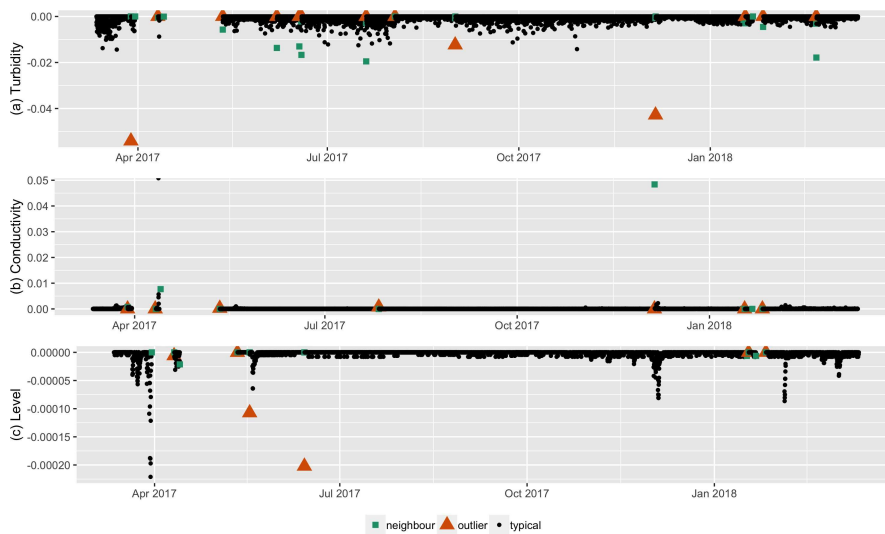

**Figure 10:** *Transformed series (one sided derivatives) of turbidity (NTU), conductivity (μS/cm) and river level (m) measured by in situ sensors at Pioneer River. In each plot, outliers determined by water-quality experts are shown in red, while typical points are shown in black.*

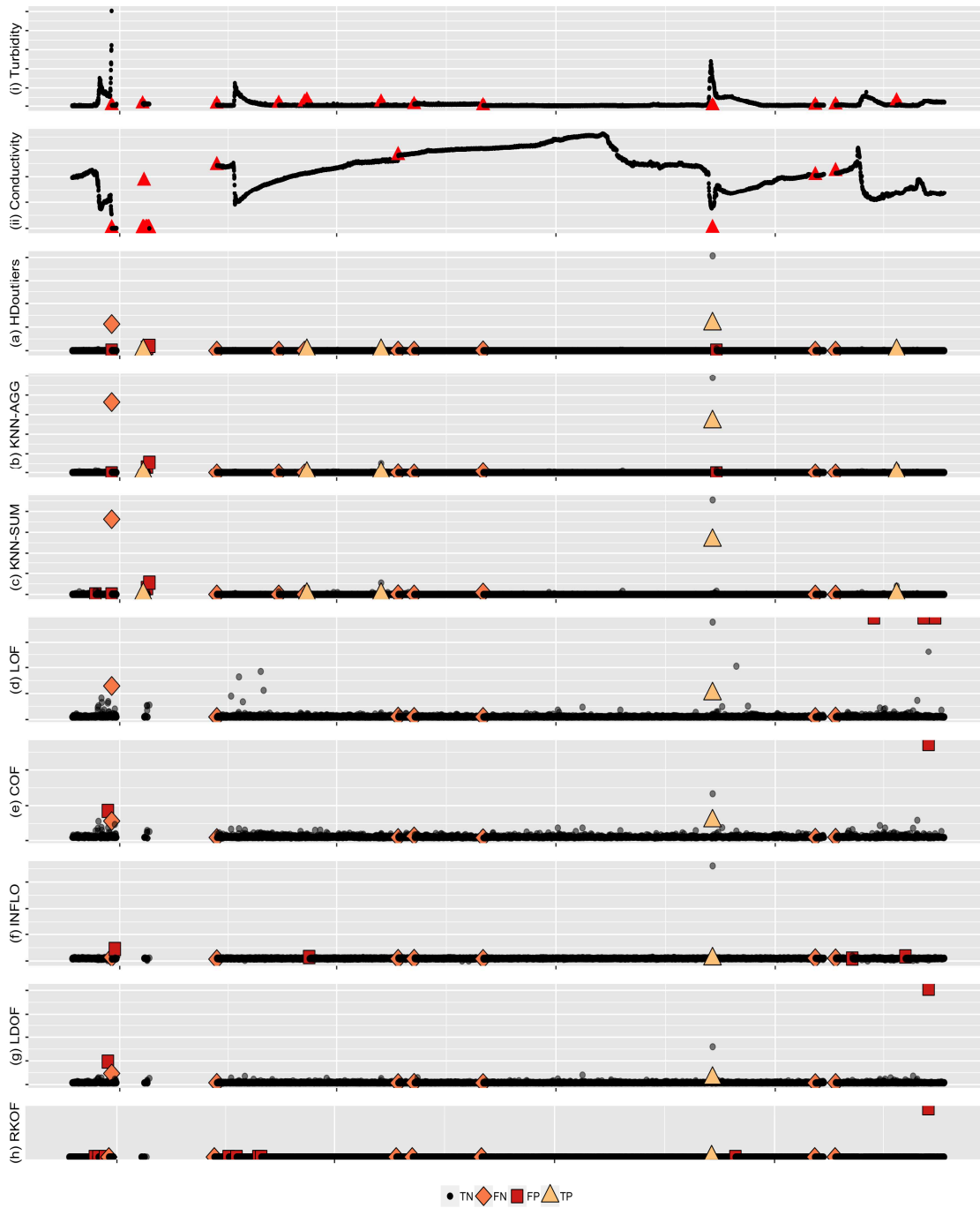

**Figure 11:** *Classification of outlier scores produced from different algorithms as true negatives (TN), true positives (TP), false negatives (FN), false positives (FP). The top two panels (i and ii) correspond to the original series (turbidity and conductivity) measured by in situ sensors at Pioneer River. The target outliers (detected by water-quality experts) are shown in red, while typical points are shown in black. The remaining panels (a–h) give outlier scores produced by different outlier detection algorithms for high dimensional data when applied to the transformed series (one sided derivative) of the two variables: turbidity and conductivity.*

**Table 3:** *Performance metrics of outlier detection algorithms performed on multivariate water-quality time series data (T, turbidity; C, conductivity; L, river level) from in situ sensors at Pioneer River, arranged in descending order of OP values. See Sections 2.7-8 for performance metric codes and details.*

| i | Variables | Transformation | Method | TN | FN | FP | TP | Accuracy | ER | GM | OP | PPV | NPV | min_t | mu_t | max_t |
|---|---|---|---|---|---|---|---|---|---|---|---|---|---|---|---|---|
| 1 | T-C | One sided Derivative | KNN-AGG | 6227 | 10 | 4 | 39 | 0.9978 | 0.0022 | 492.8012 | 0.8845 | 0.9070 | 0.9984 | 443.47 | 478.81 | 564.63 |
| 2 | T-C | One sided Derivative | KNN-SUM | 6227 | 10 | 4 | 39 | 0.9978 | 0.0022 | 492.8012 | 0.8845 | 0.9070 | 0.9984 | 209.56 | 222.25 | 325.52 |
| 3 | T-C | One sided Derivative | HDoutliers | 6226 | 10 | 5 | 39 | 0.9976 | 0.0024 | 492.7616 | 0.8844 | 0.8864 | 0.9984 | 128.02 | 136.50 | 257.66 |
| 4 | T-C-L | One sided Derivative | KNN-AGG | 6225 | 12 | 4 | 39 | 0.9975 | 0.0025 | 492.7220 | 0.8644 | 0.9070 | 0.9981 | 437.43 | 465.20 | 541.55 |
| 5 | T-C-L | One sided Derivative | KNN-SUM | 6225 | 12 | 4 | 39 | 0.9975 | 0.0025 | 492.7220 | 0.8644 | 0.9070 | 0.9981 | 195.65 | 214.48 | 297.85 |
| 6 | T-C | First Derivative | HDoutliers | 6229 | 12 | 2 | 37 | 0.9978 | 0.0022 | 480.0760 | 0.8584 | 0.9487 | 0.9981 | 169.56 | 181.98 | 272.29 |
| 7 | T-C | First Derivative | KNN-AGG | 6229 | 12 | 2 | 37 | 0.9978 | 0.0022 | 480.0760 | 0.8584 | 0.9487 | 0.9981 | 449.46 | 488.52 | 588.18 |
| 8 | T-C | First Derivative | KNN-SUM | 6229 | 12 | 2 | 37 | 0.9978 | 0.0022 | 480.0760 | 0.8584 | 0.9487 | 0.9981 | 212.09 | 225.31 | 325.85 |
| 9 | T-C | First Derivative | INFLO | 6225 | 12 | 6 | 37 | 0.9971 | 0.0029 | 479.9219 | 0.8581 | 0.8605 | 0.9981 | 1452.11 | 1525.04 | 1613.16 |
| 10 | T-C | First Derivative | RKOF | 6224 | 12 | 7 | 37 | 0.9970 | 0.0030 | 479.8833 | 0.8580 | 0.8409 | 0.9981 | 400.16 | 430.39 | 523.93 |
| 11 | T-C-L | First Derivative | RKOF | 6211 | 13 | 18 | 38 | 0.9951 | 0.0049 | 485.8168 | 0.8504 | 0.6786 | 0.9979 | 396.94 | 425.92 | 503.39 |
| 12 | T-C | First Derivative | COF | 6230 | 13 | 1 | 36 | 0.9978 | 0.0022 | 473.5821 | 0.8449 | 0.9730 | 0.9979 | 7812.66 | 7908.18 | 8453.26 |
| 13 | T-C | First Derivative | LDOF | 6230 | 13 | 1 | 36 | 0.9978 | 0.0022 | 473.5821 | 0.8449 | 0.9730 | 0.9979 | 23241.02 | 23435.72 | 24522.09 |
| 14 | T-C | One sided Derivative | COF | 6229 | 13 | 2 | 36 | 0.9976 | 0.0024 | 473.5441 | 0.8448 | 0.9474 | 0.9979 | 7393.61 | 7505.50 | 8037.09 |
| 15 | T-C | First Derivative | LOF | 6228 | 13 | 3 | 36 | 0.9975 | 0.0025 | 473.5061 | 0.8447 | 0.9231 | 0.9979 | 562.59 | 594.40 | 668.26 |
| 16 | T-C | One sided Derivative | INFLO | 6227 | 13 | 4 | 36 | 0.9973 | 0.0027 | 473.4681 | 0.8447 | 0.9000 | 0.9979 | 1488.85 | 1559.88 | 1633.10 |
| 17 | T-C | One sided Derivative | LDOF | 6228 | 13 | 3 | 36 | 0.9975 | 0.0025 | 473.5061 | 0.8447 | 0.9231 | 0.9979 | 22802.15 | 22986.04 | 23561.27 |
| 18 | T-C | One sided Derivative | LOF | 6228 | 13 | 3 | 36 | 0.9975 | 0.0025 | 473.5061 | 0.8447 | 0.9231 | 0.9979 | 581.94 | 596.93 | 682.58 |
| 19 | T-C | Original Series | INFLO | 6227 | 13 | 4 | 36 | 0.9973 | 0.0027 | 473.4681 | 0.8447 | 0.9000 | 0.9979 | 1405.56 | 1498.49 | 1578.23 |
| 20 | T-C | One sided Derivative | RKOF | 6219 | 13 | 12 | 36 | 0.9960 | 0.0040 | 473.1638 | 0.8440 | 0.7500 | 0.9979 | 388.58 | 419.66 | 510.38 |
| 21 | T-C-L | First Derivative | KNN-AGG | 6227 | 14 | 2 | 37 | 0.9975 | 0.0025 | 479.9990 | 0.8385 | 0.9487 | 0.9978 | 460.76 | 477.95 | 570.26 |
| 22 | T-C-L | First Derivative | KNN-SUM | 6227 | 14 | 2 | 37 | 0.9975 | 0.0025 | 479.9990 | 0.8385 | 0.9487 | 0.9978 | 201.55 | 220.02 | 292.20 |
| 23 | T-C | Original Series | COF | 6231 | 14 | 0 | 35 | 0.9978 | 0.0022 | 466.9957 | 0.8311 | 1.0000 | 0.9978 | 7518.53 | 7617.61 | 8501.09 |
| 24 | T-C | Original Series | LDOF | 6231 | 14 | 0 | 35 | 0.9978 | 0.0022 | 466.9957 | 0.8311 | 1.0000 | 0.9978 | 22770.85 | 22910.42 | 23857.08 |
| 25 | T-C | Original Series | LOF | 6231 | 14 | 0 | 35 | 0.9978 | 0.0022 | 466.9957 | 0.8311 | 1.0000 | 0.9978 | 551.19 | 579.15 | 632.60 |
| 26 | T-C | Original Series | HDoutliers | 6230 | 14 | 1 | 35 | 0.9976 | 0.0024 | 466.9582 | 0.8310 | 0.9722 | 0.9978 | 159.87 | 170.97 | 278.18 |
| 27 | T-C | Original Series | KNN-AGG | 6226 | 14 | 5 | 35 | 0.9970 | 0.0030 | 466.8083 | 0.8307 | 0.8750 | 0.9978 | 434.16 | 468.67 | 553.01 |
| 28 | T-C | Original Series | KNN-SUM | 6226 | 14 | 5 | 35 | 0.9970 | 0.0030 | 466.8083 | 0.8307 | 0.8750 | 0.9978 | 192.89 | 211.57 | 305.80 |
| 29 | T-C | Original Series | RKOF | 6222 | 14 | 9 | 35 | 0.9963 | 0.0037 | 466.6583 | 0.8304 | 0.7955 | 0.9978 | 373.60 | 401.90 | 475.38 |
| 30 | T-C-L | First Derivative | COF | 6228 | 15 | 1 | 36 | 0.9975 | 0.0025 | 473.5061 | 0.8251 | 0.9730 | 0.9976 | 7823.02 | 7910.74 | 8344.67 |
| 31 | T-C-L | First Derivative | LDOF | 6228 | 15 | 1 | 36 | 0.9975 | 0.0025 | 473.5061 | 0.8251 | 0.9730 | 0.9976 | 23220.06 | 23357.72 | 23878.30 |
| 32 | T-C-L | One sided Derivative | HDoutliers | 6228 | 15 | 1 | 36 | 0.9975 | 0.0025 | 473.5061 | 0.8251 | 0.9730 | 0.9976 | 125.71 | 131.88 | 206.13 |
| 33 | T-C-L | First Derivative | HDoutliers | 6227 | 15 | 2 | 36 | 0.9973 | 0.0027 | 473.4681 | 0.8250 | 0.9474 | 0.9976 | 157.28 | 167.14 | 244.33 |
| 34 | T-C-L | One sided Derivative | INFLO | 6226 | 15 | 3 | 36 | 0.9971 | 0.0029 | 473.4300 | 0.8250 | 0.9231 | 0.9976 | 1383.45 | 1418.79 | 1477.43 |
| 35 | T-C-L | One sided Derivative | COF | 6227 | 15 | 2 | 36 | 0.9973 | 0.0027 | 473.4681 | 0.8250 | 0.9474 | 0.9976 | 7414.60 | 7497.50 | 7899.60 |
| 36 | T-C-L | One sided Derivative | LDOF | 6227 | 15 | 2 | 36 | 0.9973 | 0.0027 | 473.4681 | 0.8250 | 0.9474 | 0.9976 | 22756.83 | 23090.69 | 23941.13 |
| 37 | T-C-L | One sided Derivative | RKOF | 6214 | 15 | 15 | 36 | 0.9952 | 0.0048 | 472.9736 | 0.8240 | 0.7059 | 0.9976 | 390.47 | 422.10 | 490.33 |
| 38 | T-C-L | First Derivative | INFLO | 6229 | 16 | 0 | 35 | 0.9975 | 0.0025 | 466.9208 | 0.8114 | 1.0000 | 0.9974 | 1344.71 | 1398.29 | 1456.92 |
| 39 | T-C-L | First Derivative | LOF | 6229 | 16 | 0 | 35 | 0.9975 | 0.0025 | 466.9208 | 0.8114 | 1.0000 | 0.9974 | 585.58 | 600.73 | 688.10 |
| 40 | T-C-L | Original Series | INFLO | 6229 | 16 | 0 | 35 | 0.9975 | 0.0025 | 466.9208 | 0.8114 | 1.0000 | 0.9974 | 1329.91 | 1372.84 | 1415.02 |
| 41 | T-C-L | Original Series | COF | 6229 | 16 | 0 | 35 | 0.9975 | 0.0025 | 466.9208 | 0.8114 | 1.0000 | 0.9974 | 7596.55 | 7707.19 | 8357.82 |
| 42 | T-C-L | Original Series | LDOF | 6229 | 16 | 0 | 35 | 0.9975 | 0.0025 | 466.9208 | 0.8114 | 1.0000 | 0.9974 | 22897.71 | 127337.08 | 10458495.99 |
| 43 | T-C-L | Original Series | LOF | 6229 | 16 | 0 | 35 | 0.9975 | 0.0025 | 466.9208 | 0.8114 | 1.0000 | 0.9974 | 549.48 | 580.90 | 646.90 |
| 44 | T-C-L | Original Series | HDoutliers | 6228 | 16 | 1 | 35 | 0.9973 | 0.0027 | 466.8833 | 0.8113 | 0.9722 | 0.9974 | 152.85 | 163.04 | 231.16 |
| 45 | T-C-L | Original Series | KNN-AGG | 6224 | 16 | 5 | 35 | 0.9967 | 0.0033 | 466.7333 | 0.8110 | 0.8750 | 0.9974 | 439.25 | 456.27 | 534.21 |
| 46 | T-C-L | Original Series | KNN-SUM | 6224 | 16 | 5 | 35 | 0.9967 | 0.0033 | 466.7333 | 0.8110 | 0.8750 | 0.9974 | 186.52 | 201.42 | 269.64 |
| 47 | T-C-L | One sided Derivative | LOF | 6223 | 16 | 6 | 35 | 0.9965 | 0.0035 | 466.6958 | 0.8109 | 0.8537 | 0.9974 | 583.40 | 596.14 | 672.08 |
| 48 | T-C-L | Original Series | RKOF | 6217 | 16 | 12 | 35 | 0.9955 | 0.0045 | 466.4708 | 0.8104 | 0.7447 | 0.9974 | 368.34 | 406.82 | 497.17 |

(0.9978), lowest error rates (0.0022), highest geometric means (492.8012), highest OP (0.8845) and highest NPV (0.9984). Despite the challenge given by the small spikes which could not be clearly separated from the typical points, KNN-AGG, KNN-SUM and HDoutliers with one sided derivatives of turbidity and conductivity still detected some of those points as outliers while maintaining low false negative and false positive rates. Similar to Sandy Creek, HDoutliers (< 200 milliseconds on average) and KNN-SUM (< 230 milliseconds on average) demonstrated the highest computational efficiency for the data obtained from Pioneer River.

# 4 Discussion

We proposed a new framework for the detection of outliers in water-quality data from *in situ* sensors, where outliers were specifically defined as due to technical errors that make the data unreliable and untrustworthy. We showed that our proposed framework, with carefully selected data transformation methods derived from data features, can greatly assist in increasing the performance of a range of existing outlier detection algorithms. Our framework and analysis using data obtained from *in situ* sensors positioned at two study sites, Sandy Creek and Pioneer River, performed well with outlier types such as sudden isolated spikes, sudden isolated drops, and level shifts, while maintaining low false detection rates. As an unsupervised framework, our approach can be easily extended to other water-quality variables, other sites and also to other outlier detection tasks in other application domains. The only requirement is to select suitable transformation methods according to the data features that differentiate the outlying instances from the typical behaviors of a given system.

Studies have shown that transforming variables affects densities, relative distances and orientation of points within the data space and therefore can improve the ability to perceive patterns in the data which are not clearly visible in the original data space (Dang & Wilkinson 2014). This was the case in our study, where no clear separation was visible between outliers and typical data points in the original data space, but a clear separation was obtained between the two sets of points once the one-sided derivative transformation was applied to the original series. Having this type of a separation between outliers and typical points is important before applying unsupervised outlier detection algorithms for high dimensional data because the methods are usually based on the definition of outliers in terms of distance or density (Talagala et al. 2018). Most of the outlier detection algorithms (KNN-SUM, KNN-AGG, HDoutliers, COF, LOF and INFLO) performed least well with the untransformed original series, demonstrating how data transformation methods can assist in improving the ability of outlier detection algorithms while maintaining low false detection rates.

Although outlying points were clearly separated from their majority, which corresponded to the typical behaviors, the individual outliers were not isolated and were surrounded by the other outlying points. Because HDoutliers has the additional requirement of isolation in addition to clear separation between outlying points and typical points, it performed poorly in comparison to the two KNN distance-based algorithms (KNN-AGG and KNN-SUM) which are not restricted to the single most nearest neighbor (Talagala et al. 2018). For the current work $k$ was set to 10, the maximum default value of $k$ in Madsen (2018), because too large a value of $k$ could skew the focus towards global outliers (points that deviates significantly from the rest of the data set) alone (Zhang, Hutter & Jin 2009) and make the algorithms computationally inefficient. On the other hand, too small a value of $k$ could incorporate an additional assumption of isolation into the algorithm, as in HDoutliers algorithm where $k = 1$. Among the analysis using transformed series, LOF with the first derivative transformation performed the least well, which could also be due to its additional assumption of isolation (Tang et al. 2002).

We took the correlation structure between the variables into account when detecting outliers as some were apparent only in the high dimensional space but not when each variable was considered independently (Ben-Gal 2005). A negative relationship was observed between conductivity and turbidity and also between conductivity and level for the Sandy Creek data. However, for Pioneer River, no clear relationship was observed between level and the remaining two variables, turbidity and conductivity. This could be one reason why the variable combination with river level gave poor results for the Pioneer River data set, while results for other combinations were similar to those of Sandy Creek.

The one-sided derivative transformation outperformed the derivative transformation. This was expected because in an occurrence of a sudden spike or isolated drop the first derivative assigns high values to two consecutive points, the actual outlying point as well as the neighboring point, and therefore increases the false positive rate (because the neighboring points that are declared to be outliers actually correspond to typical points in the original data space). Our goal was to detect suitable transformations, combinations of variables, and the algorithms for outlier score calculation for the two study sites. Results may depend on the characteristics of the time series (site and time dependent for example), and what is best for one site may not be the best for another site. Therefore care should be taken to select transformations most suitable for the problem in hand. According to (Dang & Wilkinson 2014), any transformation used in an analytic must be evaluated in terms of a figure of merit. For our work of detecting outliers, the figure of merit was the maximum separability of the two classes generated by outliers and typical points. However, we acknowledge that the set of transformations that we used for this work was relatively limited and influenced by the data obtained from the two study sites. Therefore, the set of transformations we considered (Table 1) should be viewed only as an illustration of our proposed framework for detecting outliers. We expect that the set of transformations will expand over time as the framework is used for other data from other study sites and for applications from other fields.

Not surprisingly, HDoutliers algorithm required the least computational time given the outlying score calculation only involves searching for the single most nearest neighbors of each test point (Wilkinson 2018). The mean computational time of KNN-AGG was twice as high as that of KNN-SUM because the KNN-AGG algorithm has the additional requirement of calculating weights that assign nearest neighbors higher weight relative to the neighbors farther apart (Angiulli & Pizzuti 2002). LOF and its extensions (INFLO, COF and LDOF) required the most computational time; all four algorithms involve a two step searching mechanism at each test point when calculating the corresponding outlying score. This means that at each test point each algorithm searches its *k* nearest neighbors as well those of the detected nearest neighbors for the outlier score calculation (Breunig et al. 2000; Tang et al. 2002; Jin et al. 2006; Zhang, Hutter & Jin 2009).

We hope to expand the ability of our proposed framework so that it can detect other outlier types not previously targeted but commonly observed in water quality data (Leigh et al. 2018). We also hope to extend our multivariate outlier detection framework into space and time so that it can deal with the spatio-temporal correlation structure along branching river networks. For the current work we selected transformation methods mainly focusing on abrupt changes in the water-quality data. Further work is recommended to explore the ability of other transformations to capture other anomaly types such as high and low variability and drift. One possibility is to consider the residuals at each point, defined as the difference between the actual values and the fitted values (similar to Schwarz (2008)) or the difference between the actual values and the predicted values (similar to Hill & Minsker (2006))), as a transformation and apply outlier detection algorithms to high dimensional space defined by the residuals. Here the challenge will be to identify the appropriate curve fitting and prediction models to generate the residual series. In this way, continuous subsequences of high values could correspond to other kinds of technical outliers such as high variability or drift.

## Acknowledgments and Data

Funding for this project was provided by the Queensland Department of Environment and Science (DES) and the ARC Centre of Excellence for Mathematical and Statistical Frontiers (ACEMS). The authors would like to acknowledge the Queensland Department of Environment and Science; in particular, the Great Barrier Reef Catchment Loads Monitoring Program for the data, and the staff from Water Quality and Investigations for their input. We thank Ryan S. Turner and Erin E. Peterson for several valuable discussions regarding project requirements and water quality characteristics. The datasets used for this article are available in the open source R package `oddwater` (Talagala & Hyndman 2018).

# Appendix

We considered the following outlier scoring techniques for the framework presented in this paper. The proposed framework can be easily updated with other unsupervised outlier scoring techniques.

### HDoutliers algorithm

The HDoutliers algorithm (Wilkinson 2018) is an unsupervised outlier detection algorithm that searches for outliers in high dimensional data assuming there is a large distance between outliers and the typical data. Nearest neighbor distances between points are used to detect outliers. However, variables with large variance can bring disproportional influence on Euclidean distance calculation. Therefore, the columns of the data sets are first normalized such that the data are bounded by the unit hyper-cube. To deal with data sets with a large number of observations, the Leader algorithm (Hartigan 1975) is used to form several clusters of points in one pass through the data set. This is done with the aim of handling micro clusters (Goldstein & Uchida 2016). After forming small mutually exclusive clusters, a representative member is selected from each cluster. The nearest neighbor distances are then calculated for the selected representative members. Using extreme value theory, an outlying threshold is calculated to differentiate outliers from typical points (see Section 2.6 for detail).

### KNN-AGG and KNN-SUM algorithms

The HDoutliers algorithm uses only nearest neighbor distances to detect outliers under the assumption that any outlying point (or outlying clusters of points) present in the data set is isolated. For example, if there are two outlying points (or two outlying clusters of points) that are close to one another, but are far away from the rest of the valid data points, then the two outlying points (or two outlying clusters of points) become nearest neighbors to one another and give a small nearest neighbor distance for each outlying point (or each outlying cluster of points). Because the HDoutlier algorithm is dependent on the nearest neighbor distances, and the two outlying points (or two outlying clusters of points) do not show any significant deviation from other typical points with respect to nearest neighbor distance, the HDoutliers algorithm now fails to detect these points as outliers.

Following Angiulli & Pizzuti (2002), Madsen (2018) proposed two algorithms: aggregated $k$-nearest neighbor distance (KNN-AGG); and sum of distance of $k$-nearest neighbors (KNN-SUM) to overcome this limitation by incorporating $k$ nearest neighbor distances for the outlier score calculation. The algorithms start by calculating the $k$ nearest neighbor distances for each point. The *k*-dimensional tree (kd-tree) algorithm (Bentley 1975) is used to identify the *k* nearest neighbors of each point in a fast and efficient manner. A weight is then calculated using the *k* nearest neighbor distances and the observations are ranked such that outliers are those points having the largest weights. For KNN-SUM, the weight is calculated by taking the summation of the distances to the *k* nearest neighbors. For KNN-AGG, the weight is calculated by taking a weighted sum of distances to *k* nearest neighbors, assigning nearest neighbors higher weight relative to the neighbors father apart.

### LOF algorithm

The Local Outlier Factor (LOF) algorithm (Breunig et al. 2000) calculates an outlier score based on how isolated a point is with respect to its surrounding neighbors. Data points with a lower density than their surrounding points are identified as outliers. The local reachable density of a point is calculated by taking the inverse of the average readability distance based on the *k* (user defined) nearest neighbors. This density is then compared with the density of the corresponding nearest neighbors by taking the average of the ratio of the local reachability density of a given point and that of its nearest neighbors.

### COF algorithm

One limitation of LOF is that it assumes that the outlying points are isolated and therefore fails to detect outlying clusters of points that share few outlying neighbors if *k* is not appropriately selected (Tang et al. 2002). This is known as a masking problem (Hadi 1992), i.e. LOF assumes both low density and isolation to detect outliers. However, isolation can imply low density, but the reverse does not always hold. In general, low density outliers result from deviation from a high density region and an isolated outlier results from deviation from a connected dense pattern. Tang et al. (2002) addressed this problem by introducing a Connectivity-based Outlier Factor (COF) that compares the average chaining distances between points subject to outlier scoring and the average of that of its neighboring to their own *k*-distance neighbors.

### INFLO algorithm

Detection of outliers is challenging when data sets contain adjacent multiple clusters with different density distributions (Jin et al. 2006). For example, if a point from a sparse cluster is close to a dense cluster, this could be misclassified as an outlier with respect to the local neighborhood as the density of the point could be derived from the dense cluster instead of the sparse cluster itself. This is another limitation of LOF (Breunig et al. 2000). The Influenced Outlierness (INFLO) algorithm (Jin et al. 2006) overcomes this problem by considering both the *k* nearest neighbors (KNNs) and reverse nearest neighbors (RNNs), which allows it to obtain a better estimation of the neighborhood's density distribution. The RNNs of an object, *p* for example, are essentially the objects that have *p* as one of their *k* nearest neighbors. Distinguish typical points from outlying points is helpful because they have no RNNs. To reduce the expensive cost incurred by searching a large number of KNNs and RNNs, the kd-tree algorithm was used during the search process.

### LDOF algorithm

The Local Distance-based Outlier Factor (LDOF) algorithm (Zhang, Hutter & Jin 2009) also uses the relative location of a point to its nearest neighbors to determine the degree to which the point deviates from its neighborhood. LDOF computes the distance for an observation to its *k*-nearest neighbors and compares the distance with the average distances of the point's nearest neighbors. In contrast to LOF (Breunig et al. 2000), which uses local density, LDOF now uses relative distances to quantify the deviation of a point from its neighborhood system. One of the main differences between the two approaches (LDOF and LOF) is that LDOF represents the typical pattern of the data set by scattered points rather than crowded main clusters as in LOF (Zhang, Hutter & Jin 2009).

### RKOF algorithm with Gaussian kernel

According to Gao et al. (2011), LOF is not accurate enough to detect outliers in complex and large data sets. Furthermore, the performance of LOF depends on the parameter *k* that determines the scale of the local neighborhood. The Robust Kernel-based Outlier Factor (RKOF) algorithm (Gao et al. 2011) tries to overcome these problems by incorporating variable kernel density estimates to address the first problem and weighted neighborhood density estimates to address the second problem

A feature-based framework for detecting technical outliers in water-quality data from in situ sensors

A feature-based framework for detecting technical outliers in water-quality data from in situ sensors